\documentclass[sigconf,screen]{acmart}
\usepackage[]{collab}
\usepackage{enumitem}
\usepackage{balance}
\usepackage{xcolor}
\usepackage{subfigure}
\usepackage{multirow}
\usepackage[most]{tcolorbox}
\usepackage{tabularx}
\usepackage{siunitx}
\usepackage{float}
\usepackage{hyperref}
\usepackage{caption}
\setcopyright{acmlicensed}
\acmDOI{10.1145/3650212.3652123}
\acmYear{2024}
\copyrightyear{2024}
\acmSubmissionID{issta24main-p448-p}
\acmISBN{979-8-4007-0612-7/24/09}
\acmConference[ISSTA '24]{Proceedings of the 33rd ACM SIGSOFT International Symposium on Software Testing and Analysis}{September 16--20, 2024}{Vienna, Austria}
\acmBooktitle{Proceedings of the 33rd ACM SIGSOFT International Symposium on Software Testing and Analysis (ISSTA '24), September 16--20, 2024, Vienna, Austria}
\received{14-DEC-2023}
\received[accepted]{2024-03-02}

\newcommand\nm{{Loghub-2.0}\xspace}
\newcommand\olddata{{Loghub-2k}\xspace}

\newcommand{\ie}{{\em i.e.},\xspace}
\newcommand{\eg}{{\em e.g.},\xspace}
\newcommand{\fixedwidth}[1]{{\ttfamily \small #1}}

\definecolor{ballblue}{rgb}{0.13, 0.67, 0.8}

\collabAuthor{yc}{gray}{Yichen Li}
\collabAuthor{jy}{ballblue}{Jinyang Liu}
\collabAuthor{zb}{teal}{Zhuangbin}
\collabAuthor{jz}{magenta}{Jiazhen Gu}
\collabAuthor{yt}{blue}{Yintong Huo}
\collabAuthor{zh}{green}{Zhihan Jiang}
\collabAuthor{jj}{purple}{Junjie}
\collabAuthor{jm}{ballblue}{Jieming Zhu}

\AtBeginDocument{%
  \providecommand\BibTeX{{%
    \normalfont B\kern-0.5em{\scshape i\kern-0.25em b}\kern-0.8em\TeX}}}

\begin{document}

\title{A Large-Scale Evaluation for Log Parsing Techniques: How Far Are We?}

\author{Zhihan Jiang}
\orcid{0009-0003-1988-6219}
\affiliation{%
  \institution{The Chinese University of Hong Kong}
  \city{Hong Kong}
  \country{China}
}
\email{zhjiang@link.cuhk.edu.hk}

\author{Jinyang Liu}
\orcid{0000-0003-0037-1912}
\affiliation{%
  \institution{The Chinese University of Hong Kong}
  \city{Hong Kong}
  \country{China}
}
\email{jyliu@cse.cuhk.edu.hk}

\author{Junjie Huang}
\orcid{0009-0004-6962-5292}
\affiliation{%
  \institution{The Chinese University of Hong Kong}
  \city{Hong Kong}
  \country{China}
}
\email{junjayhuang@outlook.com}

\author{Yichen Li}
\orcid{0009-0009-8370-644X}
\affiliation{%
  \institution{The Chinese University of Hong Kong}
  \city{Hong Kong}
  \country{China}
}
\email{ycli21@cse.cuhk.edu.hk}

\author{Yintong Huo}
\orcid{0009-0006-8798-5667}
\affiliation{%
  \institution{The Chinese University of Hong Kong}
  \city{Hong Kong}
  \country{China}
}
\email{ythuo@cse.cuhk.edu.hk}

\author{Jiazhen Gu}
\orcid{0000-0002-5831-9474}
\affiliation{%
  \institution{The Chinese University of Hong Kong}
  \city{Hong Kong}
  \country{China}
}
\email{jiazhengu@cuhk.edu.hk}

\author{Zhuangbin Chen}
\authornote{Zhuangbin Chen is the corresponding author.}
\orcid{0000-0001-5158-6716}
\affiliation{%
  \institution{Sun Yat-sen University}
  \city{Zhuhai}
  \country{China}
}
\email{chenzhb36@mail.sysu.edu.cn}

\author{Jieming Zhu}
\orcid{0000-0002-5666-8320}
\affiliation{%
  \institution{Huawei Noah's Ark Lab}
  \city{Shenzhen}
  \country{China}
}
\email{jiemingzhu@ieee.org}

\author{Michael R. Lyu}
\orcid{0000-0002-3666-5798}
\affiliation{%
  \institution{The Chinese University of Hong Kong}
  \city{Hong Kong}
  \country{China}
}
\email{lyu@cse.cuhk.edu.hk}




\renewcommand{\shortauthors}{Zhihan Jiang, et al.}

\begin{abstract}
Log data have facilitated various tasks of software development and maintenance, such as testing, debugging and diagnosing.
Due to the unstructured nature of logs, log parsing is typically required to transform log messages into structured data for automated log analysis.
Given the abundance of log parsers that employ various techniques, evaluating these tools to comprehend their characteristics and performance becomes imperative.
Loghub serves as a commonly used dataset for benchmarking log parsers, but it suffers from limited scale and representativeness, posing significant challenges for studies to comprehensively evaluate existing log parsers or develop new methods.
This limitation is particularly pronounced when assessing these log parsers for production use.
To address these limitations, we provide a new collection of annotated log datasets, denoted \nm, which can better reflect the characteristics of log data in real-world software systems.
\nm comprises 14 datasets with an average of 3.6 million log lines in each dataset.
Based on \nm, we conduct a thorough re-evaluation of 15 state-of-the-art log parsers in a more rigorous and practical setting.
Particularly, we introduce a new evaluation metric to mitigate the sensitivity of existing metrics to imbalanced data distributions.
We are also the first to investigate the granular performance of log parsers on logs that represent rare system events, offering in-depth details for software diagnosis.
Accurately parsing such logs is essential, yet it remains a challenge.
We believe this work could shed light on the evaluation and design of log parsers in practical settings, thereby facilitating their deployment in production systems.
\end{abstract}

\maketitle

\section{Introduction}

Log data records software runtime information, which is essential for developers to understand the behaviors of software systems.
The rich information encapsulated within log data empowers developers and maintainers to test programs~\cite{andrews1998testing,chen2018automated,messaoudi2021log,chen2023exploring}, identify bugs~\cite{yuan2010sherlog,fu2014developers,amar2019mining,chen2021pathidea} and diagnose softwares~\cite{nagappan2009efficiently,nagaraj2012structured,notaro2023logrule}.
In general, log messages are semi-structured textual data, generated by logging statements written by developers in the source code, \eg \fixedwidth{`logger.info(``connected to host: \{\}'', hostIp)'} in Java~\cite{zhi2019exploratory,liu2022tell,li2023exploring,li2024go}.
At runtime, the variable \fixedwidth{hostIp} may change in different executions, which can result in a sequence of log messages like \fixedwidth{`connected to host: 172.16.254.1'} and \fixedwidth{`connected to host: 172.16.254.2'}. 
Log parsing aims to convert such semi-structured log messages into structured events, which often serves as the first and foremost step to many log analysis tasks~\cite{he2021survey,le2022log,zhang2023system,ma2023automatic,liu2023scalable}.
Specifically, log parsing extracts the constant parts (\ie \textit{log templates}) and the changeable parts (\ie \textit{log parameters}) from log messages.
In the above example, the log template is \fixedwidth{`connected to host: <*>'}, and the log parameter indicates the concrete IP address of the host, \eg \fixedwidth{`172.16.254.1'}.

Traditional approaches parse logs via matching raw log messages with their respective logging statements within the source code~\cite{shang2012bridging,pecchia2015industry,schipper2019tracing,bushong2020matching}.
However, this approach is usually impractical since software source code may not always be available, \eg commercial software.
Thus, tremendous efforts have been devoted to data-driven approaches~\cite{he2017drain,yu2023brain,huo2021semparser,jiang2023llmparser}.
These parsers directly process raw log messages without access to the source code. 

Given the variety of log parsers employing different techniques, it is crucial to evaluate these tools to comprehend their characteristics and performance, providing guidance for production adoption in industry.
To this end, \citet{zhu2023loghub} released Loghub, which contains an extensive collection of log datasets generated by various systems.
However, Loghub only provides annotated parsing ground truth for 2,000 lines of logs randomly sampled from each system, denoted as \olddata, which has been extensively used to evaluate existing log parsers~\cite{zhu2019tools,khan2022guidelines} and develop new log parsers~\cite{dai2020logram,wang2022spine,li2023did,dai2023pilar}.
% To this end, \citet{zhu2019tools} released a collection of log parsing datasets, denoted as \olddata, based on Loghub.
% Loghub contains an extensive collection of log data generated by various systems, while \olddata randomly sample 2,000 lines of logs for each system and label each log with its parsing ground truth.
% Since it covers a diverse range of systems, \olddata has been extensively used to evaluate existing log parsers~\cite{zhu2019tools,khan2022guidelines} and develop new log parsers~\cite{dai2020logram,wang2022spine,li2023did,dai2023pilar}.

Even though existing log parsers, such as Drain~\cite{he2017drain}, Logram~\cite{dai2020logram} and LogPPT~\cite{le2023log}, have reported state-of-the-art results on the \olddata dataset, we observed that the parsing performance of these tools is compromised when being integrated into real-world software systems~\cite{fu2022investigating,sedki2023towards}.
Based on our experiences of deploying automated log parsing in production, we have found that existing parsers struggle to identify two types of log templates, \ie \textit{infrequent log templates} (those that occur infrequently) and \textit{parameter-intensive log templates} (those that involve many parameters). 
The former usually includes logs with severe logging levels (\eg error or fatal), which typically demand more attention due to their potential impact. 
The latter usually records the system runtime status and associated values.
These two types of log templates are crucial for downstream analysis tasks, such as anomaly detection and debugging. 
Therefore, it is essential to ensure parsing accuracy for such log templates.
However, the research results reported in previous studies~\cite{zhu2019tools,khan2022guidelines} may not necessarily apply in practical production settings, especially for these two types of log data.

This performance disparity primarily originates from three inherent limitations in existing benchmark studies.
First, the widely-used \olddata is of limited scale. 
It only encompasses 2,000 lines of log messages in each dataset, whereas real-world data often consist of millions of log lines~\cite{zhu2019tools,wang2022spine,yao2021improving}.
As a result, the \olddata may not be able to sufficiently represent the complex characteristics of log data obtained from production systems, particularly in terms of the frequency and parameter count of log templates.
Second, evaluation metrics used in existing benchmarks (\eg group accuracy, GA~\cite{zhu2019tools}) are often \textit{message-level} (\ie calculated based on the number of log messages), thus may produce misleadingly high-accuracy results.
It is because the distribution of log templates' occurrences is usually highly imbalanced in production systems~\cite{wang2022spine, khan2022guidelines}.
The evaluation results could be dominated by the majority classes of log templates (\ie those contain many log messages).
Therefore, such metrics may not be robust enough to datasets with diverse template distributions.
Third, existing studies often report the performance of log parsers in processing the entire dataset.
It is unclear how they perform when dealing with the above two types of log templates.
The lack of fine-grained evaluation can lead to a limited understanding of how well a log parser handles these specific cases in practice.

To address these limitations, we propose a new log parsing benchmark tailored to evaluate log parsers in a more rigorous and practical setting.
Specifically, 
(1) On top of the raw Loghub logs~\cite{
he2020loghub}, we build a new version of large-scale annotated log datasets for log parsing, denoted as \nm. 
The annotation is conducted adhering to a rigorous framework, which can significantly reduce manual efforts through log grouping and template matching.
% (1) we create a large-scale collection of annotated log datasets for log parsing, denoted as \nm.
% It is built upon the raw Loghub logs while adhering to a rigorous annotation framework, which can significantly reduce manual efforts through log grouping and template matching.
\nm aims to reflect the scale and distribution of log data observed in real-world scenarios.
In detail, \nm contains 14 datasets from various software systems. 
Each dataset contains \textit{3.6 million} lines of log messages on average.
Each log line has been manually annotated with its corresponding log template and parameter(s).
(2) We propose a more comprehensive benchmarking protocol to evaluate existing log parsers. 
The protocol includes a new \textit{template-level} metric, \ie F1-score of Group Accuracy (FGA), to mitigate the sensitivity of the message-level metrics (\eg GA) to imbalanced data.
Moreover, we make the first step to investigate the performance of log parsers on log templates with different frequencies and parameter counts, providing an essential reference regarding how well they would perform in production environments.
(3) We conduct an extensive re-evaluation of \textit{15} log parsers, including \textit{13} statistic-based and \textit{2} semantic-based log parsers, on \nm using the proposed benchmarking protocol.
Our study provides researchers and practitioners with a more practical perspective on understanding the characteristics of these parsers.
We summarize the key findings from the evaluation as follows.

\textbf{Key Findings.}
(1) Compared with \olddata, \nm exhibits more realistic data characteristics, especially in the context of log template frequencies and parameter counts.
(2) All existing parsers demonstrate a significant degradation in performance on \nm compared to the \olddata, with a greater degree of variance.
This shows that our proposed datasets and benchmarking protocol can reveal the performance of log parsers under more complex and diverse conditions.
(3) Achieving high overall performance on the entire datasets does not necessarily guarantee effective parsing of infrequent and parameter-intensive logs, which often deserve more attention in system maintenance.
Thus, a comprehensive evaluation should consider different types of logs to ensure robust and reliable performance in practice.
(4) \textit{9} out of \textit{15} parsers fail to process all the datasets in \nm within a reasonable time frame, highlighting the importance of improving parsing efficiency, especially for production deployment.
% This underscores the need for more attention to parsing efficiency, particularly when it is intended for production deployment.

The main contributions of this paper are summarized as follows:
\begin{itemize}[leftmargin=*, topsep=0pt]
    \item We propose a new collection of large-scale datasets for evaluating log parsing techniques, referred to as \nm.
    This collection comprises 14 datasets, each with an average of 3.6 million log lines. 
    The parsing labels of the log messages are manually annotated through a rigorous annotation framework, which ensures the efficiency and accuracy of the labeling process.
    This is a significant extension of the existing widely-used \olddata, which contains only 2,000 lines of log messages per dataset.
    
    \item We propose a more comprehensive benchmarking protocol for log parsers, which emphasizes assessing parsing accuracy on logs with different characteristics. Moreover, a new template-level metric, \ie FGA, is proposed to address the sensitivity of existing metrics to imbalanced data. 
    
    \item We re-evaluate 15 state-of-the-art log parsers by our benchmarking protocol and derive seven interesting findings, which could shed light on the design and evaluation of log parsers in a more practical setting. To benefit future research, we make datasets, source code, and experimental results publicly available~\cite{repo}.
\end{itemize}

%%%%%%%
%%%%%%%

%%%%%%%
%%%%%%%

\vspace{-4pt}
\section{Background and Motivation}

In this section, we first briefly introduce existing log parsers in the literature.
Then, we talk about the observations that motivate us to revisit existing log parsing studies.

\subsection{Existing Log Parsing Techniques}
Many log parsing approaches have been proposed in the literature, mainly classified into the following four categories.

\noindent
\textbf{Frequency-based Parsing}
This type of methods~\cite{vaarandi2003data,nagappan2010abstracting,vaarandi2015logcluster,dai2020logram} is founded on the intuition that tokens, which frequently occur within a specific log dataset, generally represent the static elements of those logs.
Consequently, the extraction of frequent patterns provides a straightforward approach for automated log parsing.
In detail, these log parsers first traverse the provided log dataset to construct frequent itemsets.
Subsequently, these itemsets are utilized to derive the corresponding log template for log messages.

\noindent
\textbf{Similarity-based Parsing}
These log parsers~\cite{fu2009LKE,tang2011logsig,mizutani2013SHISO,hamooni2016logmine,shima2016LenMa} conceptualize log parsing as clustering logs into distinct clusters predicated on their similarity, and logs in each cluster share the same log template.
Various methods employ different clustering algorithms (\eg hierarchical clustering, density-based clustering) and definitions of similarity.
Following the clustering process, log templates can be derived by extracting the common tokens from the logs within each respective cluster.

\noindent
\textbf{Heuristics-based Parsing}
Another category of log parsers~\cite{jiang2008abstracting,makanju2009clustering,du2016spell,he2017drain,messaoudi2018search,yu2023brain} employs a diverse range of heuristic algorithms or data structures, such as the longest common subsequence-based approach, parsing trees, evolutionary algorithm, among others.
These log parsers are designed to leverage the unique characteristics of log data to distinguish the templates and parameters in log messages.

\noindent
\textbf{Semantic-based Parsing}
In recent years, numerous parsers~\cite{huo2021semparser,liu2022uniparser,li2023did,le2023log} have employed deep neural networks to understand the semantic meaning of logs, thereby improving parsing accuracy. In detail, these log parsers employ supervised methodologies, utilizing models such as bidirectional long short-term memory or pre-trained language models to learn the semantic information of log messages, thereby distinguishing between log templates and parameters through the completion of classification tasks.

\vspace{-4pt}
\subsection{Motivation}
\label{sec: motivation}

\begin{figure*}[htbp]
    \centering
    \includegraphics[width=\textwidth]{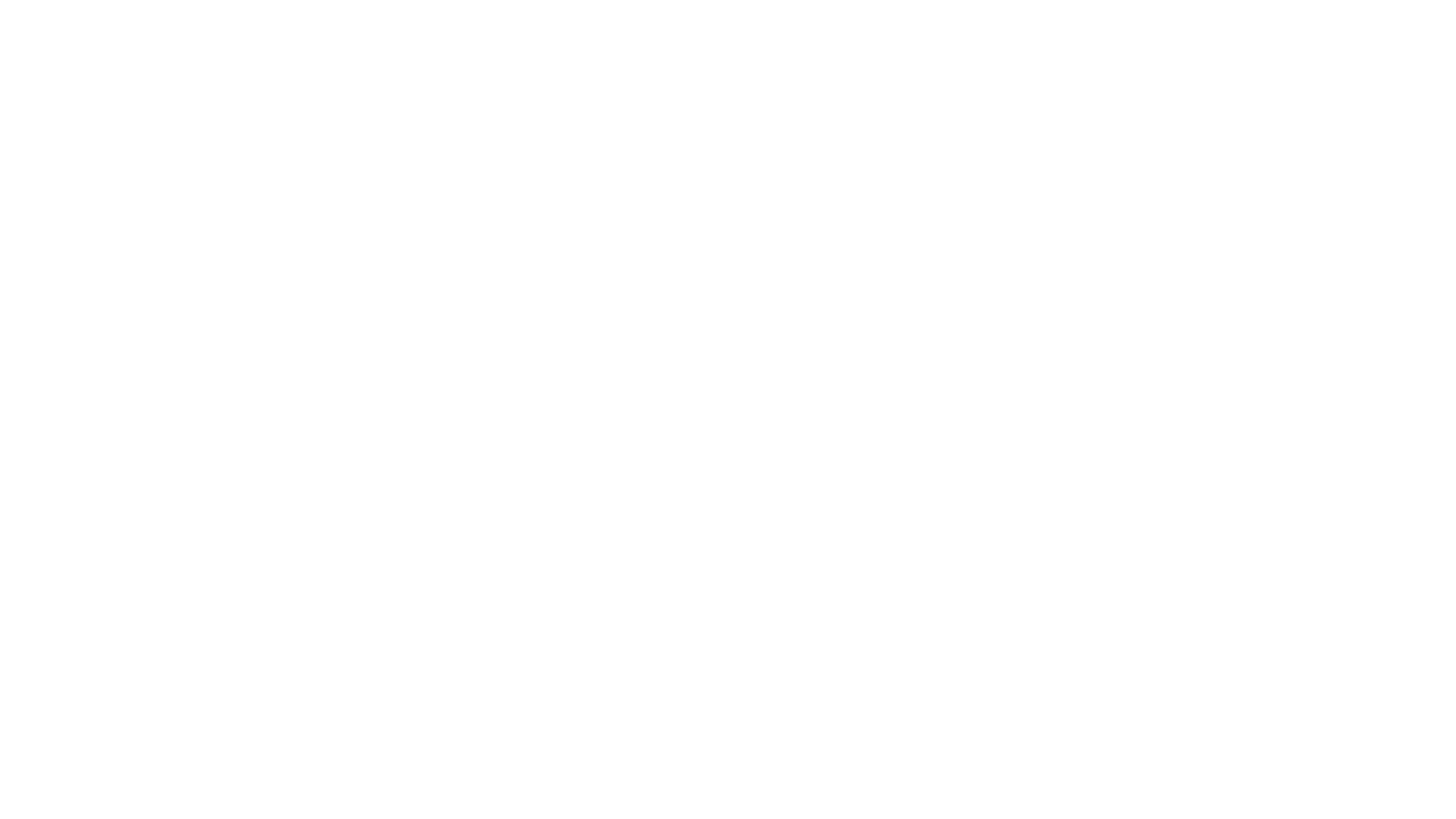}
    \caption{The overall framework of data annotation} 
    \label{fig: framework}
    \vspace{-2pt}
\end{figure*}

Given the fruitful log parsing studies, comprehensively evaluating existing log parsers is crucial in understanding their characteristics and guiding the selection of appropriate methods in practice.
\citet{zhu2019tools} proposed the first benchmark of 13 log parsers by collecting multiple log datasets from various types of systems, including distributed systems, supercomputer systems, etc.
Particularly, they randomly sampled 2,000 log messages for each dataset and manually annotated the template for these logs.
This results in the widely-used collection of log parsing datasets, \ie~\olddata.
Many new log parsing approaches~\cite{dai2020logram,wang2022spine,le2023log} also evaluate their performance on \olddata and demonstrate promising results.

Despite the advantages of the dataset, we still observe some inherent limitations associated with it.
First, a recent study~\cite{khan2022guidelines} has pinpointed multiple errors in the annotated templates, which could potentially impact the assessment of log parsers.
Therefore, they proposed several heuristic rules, such as regular expressions, to fix the incorrect templates in \olddata.
Second, we find that these log parsers demonstrate compromised effectiveness and efficiency in production deployment.
This highlights the limitations of previous benchmark studies, as they do not fully capture the comprehensive performance of log parsers, especially in practical environments.
To understand the aforementioned limitations, we have conducted a thorough investigation and identified three primary reasons:

\begin{itemize}[leftmargin=*, topsep=0pt]

\item \olddata is small in scale, with each dataset comprising only 2,000 lines of log messages.
Considering that real-world systems often produce a large volume of log data (\eg tens of gigabytes per hour~\cite{wang2022spine,he2022empirical,locke2021logassist}), \olddata may not be able to reflect the diverse and complex characteristics of log data observed in production environments.
Given the data-driven nature of most existing log parsers, their performance could be affected by the limited scale of Loghub-2k and may not generalize well to real-world scenarios with much larger and diverse log datasets.
Moreover, the annotation process of \olddata does not follow a rigorous and standardized approach, leading to potential errors and inconsistencies in the annotated templates.

\item Existing studies often lack a comprehensive set of metrics for evaluation. 
Most of them only employ message-level metrics such as group accuracy~\cite{zhu2019tools} and parsing accuracy~\cite{dai2020logram}.
Theoretically, these metrics tend to favor frequently occurring log templates.
For instance, if a simple template involves a large number of log messages, then successfully parsing this template would yield good performance, irrespective of the results on the less frequent templates.
In real-world scenarios, log data could be highly imbalanced.
For example, some systems periodically print log messages to record routine information, such as system heartbeats (``\fixedwidth{System uptime: 30 hours}'').
Such logs might not be of interest to system operators.
However, they could dominate the overall performance, masking potential errors in processing infrequent templates.

\item The current evaluation of log parsers mainly reports the overall performance on entire datasets.
This approach, however, lacks a fine-grained analysis of parsing performance for logs with different characteristics.
We have identified two critical types of logs that play an essential role in production system maintenance.
These include infrequent log templates, which represent rare system events that require particular attention, and parameter-intensive log templates, which provide informative details about system status and the associated entities.
Investigating the performance on these logs helps understand the actual effectiveness of log parsers in real-world applications.
\end{itemize}

To address these limitations, we propose conducting a new benchmark study for existing log parsers in a more rigorous and practical setting.
This entails the creation of a more diverse collection of log parsing datasets that are substantially larger in scale, as well as the design of a more comprehensive benchmarking protocol.

\vspace{-5pt}
\section{Dataset Construction}
\label{sec:data_annotation}

In this section, we introduce the construction process of the dataset collection, which intends to reflect the scale and characteristics of real-world log data and thus enables an accurate assessment of log parsers' capabilities in practical scenarios.
Particularly, we propose a rigorous annotation framework designed to ensure the efficiency, accuracy and consistency of the labeling procedure.

\vspace{-5pt}
\subsection{Overview}

To construct the datasets, we select 14 log datasets in Loghub~\cite{he2020loghub} that span different types of systems, including distributed systems, supercomputers, and operating systems.
Although these datasets are collected from various types of systems on a large scale, they lack the essential labels for log parsing assessment.
Thus, a necessary preliminary step in our study is annotating these datasets.

The annotation process is carried out by a team of five skilled data annotators. 
This team consists of three Ph.D. students with a minimum of two years of experience in system maintenance research, alongside two industry engineers, both of whom have at least five years of experience in software development and management.
Given the immense size of each dataset (\eg millions of log messages), manual labeling for each log message is infeasible.
Therefore, we design a rigorous annotation framework to assist the annotation process, which guarantees both labeling efficiency and accuracy through log grouping and template matching.

Fig.~\ref{fig: framework} shows the overview of the annotation framework, which includes five steps: \textit{preprocessing}, \textit{log grouping}, \textit{template annotation}, \textit{log-template matching}, and \textit{template refinement}.
To begin, we first preprocess the raw logs to obtain meaningful log contents.
Then, we apply a hierarchical approach to coarsely partition the logs into distinct groups.
The logs sharing the same template are highly likely to be divided into the same group, facilitating efficient annotation procedures.
Within each group, all annotators carefully identify all log templates.
{In this process, we arrange log messages within each group in lexicographical order to place similar log messages together, which enables us to quickly annotate all log templates instead of labeling each log message.}
After the annotation, we employ regular expressions to construct the matching between log messages and the labeled log templates.
If any log messages remain unmatched, we review and rectify the templates, subsequently repeating the matching process until all log messages are matched.
Finally, we conduct template refinement to calibrate the results of all annotators to ensure the accuracy and uniformity of the annotations across all annotators and datasets.
The details of each step are explained as follows.

\vspace{-5pt}
\subsection{Preprocessing}

Following previous work~\cite{he2017drain,zhu2019tools,wang2022spine}, we first apply predefined regular expressions to extract different fields of log messages.
Typical fields include timestamp, logging level, component, and content.
We then undertake a cleaning process for the logs. This process specifically targets logs whose content does not include any alphabetical characters, such as logs comprised solely of numerical figures or punctuation marks. Such logs are cleaned due to their lack of parseable content. 
We also remove log lines with duplicate content temporarily to reduce manual efforts in the following steps.

\vspace{-5pt}
\subsection{Log Grouping}
After the preprocessing stage, we are still facing a substantial number of log messages (\eg millions of log messages in the HDFS dataset), making it unfeasible to manually annotate each message.
Inspired by~\cite{liu2019logzipIC}, we adopt a hierarchical approach to coarsely divide log messages into multiple groups.
Our goal is to group together log messages sharing the same template, which enables us to label their template in one pass.
To this end, we first partition logs based on their logging level and component name, which are extracted during the preprocessing step.
These two properties provide a straightforward means to initially identify logs that belong to the same template~\cite{liu2019logzipIC}.
Second, we use more advanced information to group logs, \ie the most frequently occurring tokens of a log message.
Specifically, we employ delimiters such as spaces and punctuation to tokenize each log message into multiple tokens and calculate the frequency of each token in the dataset.
For each log message, we calculate the K most frequent tokens and then group together those log messages that share the common top-K frequent tokens.
The underlying rationale is that the template part of log messages remains stable, while the parameter part can dynamically change during runtime.
As a result, the most frequent K tokens in log messages can effectively serve as robust evidence to determine their belonging to the same group.
The value of K for each dataset has been determined based on their characteristics, with the value ranging from one to three.
Particularly, we maintain a collection of stop words to be excluded from the top-K frequent tokens, ensuring that these common words do not interfere with the grouping process.
In addition to the stop words provided by the Scipy library~\cite{scipy}, we have also manually added certain words to the collection, such as \fixedwidth{root}, \fixedwidth{true}, etc.
Finally, we obtain multiple coarse-grained groups of log messages where the log messages in each group share the same logging level, component, and top-K tokens.

\vspace{-8pt}
\subsection{Template Annotation}

The objective of the log grouping step is to partition the logs into coarse-grained groups, ensuring, as far as possible, that logs sharing the same template are divided into the same group.
Therefore, it is possible that log messages with different templates are grouped together.
To address this issue, we employ manual template annotation to derive ground-truth log templates from each group.

To accelerate the manual annotation process, we sort the log messages within each group in lexicographical order to place similar log messages together.
Annotators can then quickly recognize the templates of logs based on their structures and similarities.
Consequently, annotators are required to annotate only the ground-truth log templates, eliminating the need for sequential labeling of each log message.
This approach is based on the observation that the quantity of potential log templates is typically several orders of magnitude smaller than the total number of log messages~\cite{zhu2019tools,khan2022guidelines,wang2022spine}, which renders manual labeling a feasible task.

In detail, all annotators conduct the manual annotation process independently, whose individual results will be consolidated to generate the final results (to be detailed in Sec.~\ref{sec: template_refinement}).
To ensure the accuracy and consistency of the annotation, we adhere to the parameter categories proposed by \citet{li2023did} to determine whether a token is a parameter. 
We also ask the annotators to apply the same heuristic rules proposed by \citet{khan2022guidelines} to ensure a more consistent template format, \eg replace double spaces with a single space.
If a token is identified as a parameter, we will replace it with \texttt{"<*>"}, and the static parts remain unchanged to form the corresponding templates.
Since similar log messages have been grouped and sorted together, we can efficiently bypass numerous log messages that evidently share the same template when handling each group, and only the identified templates are recorded.
Finally, in the rare cases where different groups still share identical templates, we compare the templates derived from different groups, eliminate duplicate templates, and merge some templates as necessary.
This procedure can eliminate potential errors in the log grouping step, thereby ensuring the accuracy of the annotations.

\vspace{-8pt}
\subsection{Log-Template Matching}

While we generate log templates in the manual annotation step, we do not record the explicit matching between each log message and its corresponding template.
This is to avoid complicating the annotation process, and later in the possible deduplication and merging procedures, it could lead to ambiguous relations and challenges in maintaining clarity and accuracy.
Instead, we resort to the technique of regular expressions to automatically construct the matching between logs and templates.

Specifically, we convert each template into a regular expression by substituting \texttt{"<*>"} with \texttt{"(.*)"}, which enables each parameter position to match strings of any length. Subsequently, for each log message, we attempt to match it against every log template, halting when a match is found. 
Although this step requires pairwise matching between a large number of log messages and log templates, it can be completed within a reasonable time given that the number of log templates is typically much smaller than the total count of log messages (as shown in Table~\ref{tab: dataset}). We also further speed up this matching process by implementing it in a parallel manner. 

Additionally, in this matching process, one log message could match multiple templates. 
For examples, two templates $T_1$:  \fixedwidth{"auth failure; logname=<*> uid=<*> ruser=<*>"} and $T_2$: \fixedwidth{"auth failure; logname=<*> uid=<*>"} can exist in the same dataset. 
All log messages generated from template $T_1$ can be matched by $T_2$ since the last  \fixedwidth{<*>} are allowed to match multiple tokens.
To address this issue, when a log message matches multiple regular expressions, we give priority to the templates with longer static parts for annotation.
The intuition is that when two different templates are capable of matching the same log message, the template that can match more non \texttt{"<*>"} characters suggests a higher probability that this log message belongs to that particular template.
In the rare cases where the two are the same, we choose templates with fewer \texttt{"<*>"} to generate more compact and simple templates.
By applying this rule, the log messages of $T_1$ will be correctly matched with $T_1$.

In instances where specific log messages fail to match any regular expression, we will revert to the template annotation step, carefully review these log messages, make necessary template corrections, and subsequently repeat the matching process.
Ultimately, each log message should successfully match one regular expression that corresponds to its annotated template.

\vspace{-8pt}
\subsection{Template Refinement}
\label{sec: template_refinement}
The last step is template refinement, which aims to consolidate all five annotators' results and correct potential errors. 
After carefully comparing the templates from five annotators, we identify the following inconsistent cases that occur most frequently.
All discrepancies are addressed through discussions to ensure accurate and uniform annotation.

\begin{itemize}[leftmargin=*, topsep=0pt]
    \item One annotator may produce more templates than others. 
    In this case, it is possible that some of his annotated templates are too specific. For example, some variables (\eg \fixedwidth{root}/\fixedwidth{True}/\fixedwidth{temp}) are incorrectly identified as constant. We then regard such cases as parameters following~\cite{khan2022guidelines}.
    \item The same template may have different formats, \eg \fixedwidth{``1165 bytes (1.13 KB) sent''} may be labeled as \fixedwidth{``<*> bytes (<*> KB) sent''}  or \fixedwidth{``<*> bytes <*> sent''}. In this case, we chose the former one to retain the original format of the log messages.
\end{itemize}

Additionally, we quantitatively assess the annotation consistency of five annotators.
This is measured by determining the proportion of templates where the annotations of the five annotators are identical. 
The average consistency score across all datasets attains a value of $0.926$, indicating a high agreement in the annotation step.
Ultimately, all five annotators reached a consensus on all annotated templates, which is then adopted as the final annotation.

\vspace{-8pt}
\subsection{Annotation Results}
\label{sec: annotation_results}
The data annotation process finally produces a collection of large-scale log paring datasets from diverse systems, named \nm. 
The detailed statistic of \nm is presented in Table \ref{tab: dataset}. 
Compared with \olddata, the average number of annotated log messages has seen a substantial increase, escalating by a factor of 1875, from 2,000 to 3,601,187.
Furthermore, the average number of annotated log templates has increased by 204.2\%, from 81.9 to 249.1, encompassing a broader range of templates.
The large scale of the new dataset collection, \nm, enables detailed evaluations of log parsing techniques, potentially exposing their performance in more realistic and large-scale scenarios.

\begin{table}[]
\small
\caption{Statistics of \nm}
\vspace{-5pt}
\centering
\label{tab: dataset}
 \resizebox{0.49\textwidth}{!}{%
\begin{tabular}{cccccc}
\toprule
\multirow{2}{*}{\textbf{\shortstack{System}}} & \multirow{2}{*}{\textbf{\shortstack{Dataset}}} & \multirow{2}{*}{\textbf{\shortstack{\# Templates\\(\olddata)}}} & \multirow{2}{*}{\textbf{\shortstack{\# Templates\\(\nm)}}}  & \multirow{2}{*}{\textbf{\shortstack{\# Annotated Logs\\(\nm)}}} \\
 & & & & & \\
 \midrule
 \multirow{5}{*}{\shortstack{Distributed\\systems}} & Hadoop & 114  & 236 & 179,993  \\
 & HDFS & 14  & 46 & 11,167,740 \\
 & OpenStack &  43 & 48 & 207,632 \\
 & Spark & 36  & 236 & 16,075,117 \\
 & Zookeeper & 50 & 89 & 74,273 \\
\midrule
\multirow{3}{*}{\shortstack{Super-\\computer\\systems}} & BGL & 120  & 320 & 4,631,261 \\
 & HPC & 46  & 74  & 429,987  \\
 & Thunderbird & 149 & {1,241} & {16,601,745} \\
\midrule
\multirow{2}{*}{\shortstack{Operating\\systems}} &  Linux & 118 & 338 & 23,921 \\
 & Mac & 341  & {626} & {100,314} \\
\midrule
\multirow{2}{*}{\shortstack{Server\\application}} & Apache & 6 & 29 & 51,977 \\
 & OpenSSH & 27 & 38 & 638,946 \\
\midrule
\multirow{2}{*}{\shortstack{Standalone\\software}} & HealthApp & 75  & 156 & 212,394 \\ 
 & Proxifier & 8 & 11 & 21,320\\
\bottomrule
\multicolumn{2}{c}{\textbf{Average}} & \textbf{81.9} & \textbf{249.1} & \textbf{3,601,187} \\
\bottomrule
\end{tabular}
}
\vspace{-10pt}
\end{table}

\vspace{-6pt}
\section{Study Design}

In this section, we introduce the design of our benchmark study for log parsers.
Based on the large scale and diversity of \nm, we aim to gain a more in-depth understanding of the log parsers' effectiveness and suitability for real-world applications.
To this end, we first design three research questions to guide the study.
Then, we select a new set of metrics for comprehensive performance assessment, which includes a new template-level metric that we design and existing popular metrics.
Finally, we elaborate on the selected log parsers for evaluation and the experiment setup.

\vspace{-6pt}
\subsection{Research Question}

\noindent
\textbf{RQ1: What are the differences between \nm and \olddata ?}
In this RQ, we aim to explore whether there are significant differences in the characteristics of \nm and \olddata, which could potentially impact the performance of log parsers. 
Specifically, We focus on examining two important characteristics: \textit{frequencies of log templates} and \textit{parameter counts in log templates}. 

\noindent
\textbf{RQ2: How does the performance of log parsers differ when applied to \nm compared to \olddata?}
In this RQ, our focus lies in conducting a comprehensive re-evaluation of log parsers using \nm, encompassing both effectiveness and efficiency aspects. We also explore any potential limitations associated with the widely-used \olddata. To this end, we carefully compare the evaluation results obtained from \nm with those from \olddata, enabling us to draw insightful conclusions.

\noindent
\textbf{RQ3: What is the performance of log parsers on logs with varying characteristics?}
Inspired by our observations in Sec.~\ref{sec: motivation}, we investigate the performance of log parsers on logs with diverse template frequencies and parameter counts.
This is pivotal, as certain logs with distinctive characteristics may hold significant importance in production environments.
Particularly, such an evaluation becomes feasible only with the use of the labeled datasets in \nm, attributable to its large scale and diversity.

\vspace{-6pt}
\subsection{Evaluation Metrics}
\label{sec:study_metrics}

We employ two categories of metrics, \ie message-level and template-level metrics, to evaluate log parsers.
\textit{Message-level metrics} account for the quantities of messages belonging to each template, thereby favoring templates with a higher volume of log messages. 
On the other hand, \textit{template-level metrics} evenly consider each template, regardless of the number of log messages each template corresponds to.
In our benchmark protocol, we adopt two message-level metrics \ie Group Accuracy (GA) and Parsing Accuracy (PA) and two template-level metrics \ie F1-score of Group Accuracy (FGA) and the F1-score of Template Accuracy (FTA)~\cite{khan2022guidelines}. 
In particular, FGA, proposed by us, is the template-level version of GA.
Below, we elaborate on the metrics used in our study.

\vspace{-6pt}
\subsubsection{Message-Level Metrics}

Following existing studies, we utilize two popular message-level metrics, \ie GA and PA.

\noindent
\textbf{Group Accuracy (GA).}
GA is first used by \citet{zhu2019tools}, which assesses the ability to correctly group log messages belonging to the same template. It is defined as \textit{the proportion of correctly grouped log messages to the total number of log messages}. A log message is regarded as correctly grouped if and only if its template corresponds to the same set of log messages as the ground truth does.

\noindent
\textbf{Parsing Accuracy (PA).} 
PA utilized by \citet{dai2020logram} assesses the ability to correctly extract the template parts and parameter parts for each log message, which is essential for various log analysis tasks, such as anomaly detection using parameter values~\cite{jia2017logsed,du2017deeplog,khan2023impact}.
It is defined as \textit{the ratio of correctly parsed log messages over the total number of log messages}, where a log message is considered to be correctly parsed if and only if all tokens of static templates and dynamic variables are correctly identified.

\vspace{-6pt}
\subsubsection{Template-Level Metrics}
\label{sec: template-level metrics}

Despite the wide use of message-level metrics~\cite{yu2023brain,li2023did,yu2023self}, they consider the number of log messages and thus are sensitive to imbalanced templates.
For example, in a dataset where 95\% of log messages belong to only 1\% of the templates, a log parser could achieve a GA or PA of 0.95 by accurately grouping or parsing these 1\% of templates, regardless of any parsing errors for the remaining 99\% of templates.
In practice, certain infrequently occurring templates, such as error-level log messages, may hold crucial significance, while frequently appearing templates, like heartbeat messages, might be of less importance.
Thus, template-level metrics, which do not consider the number of log messages of each template, should also be incorporated to comprehensively evaluate the performance of log parsers.

\noindent
\textbf{F1-score of Group Accuracy (FGA).}
We propose FGA, which focuses on the proportion of correctly grouped templates rather than log messages.
Thus, it can be considered as calculating GA at the template level.
Specifically, FGA is the harmonic mean of PGA (Precision of Group Accuracy) and RGA (Recall of Group Accuracy).
Let $N_p$ be the number of templates that are generated by a log parser, and $N_c$ be the number of templates that are correctly parsed by the log parser. 
The correctness here has the same definition as in GA, \ie a log template is considered as correctly parsed if and only if the set of log messages belonging to this template matches the set indicated in the ground truth.
$N_g$ is the actual correct number of templates in the ground truth.
Based on these notations, we can define PGA as $\frac{N_c}{N_p}$ and RGA as $\frac{N_c}{N_g}$.
Then, we can calculate FGA as their harmonic mean, \ie $\frac{2\times PGA\times RGA}{PGA+RGA}$

\noindent
\textbf{F1-score of Template Accuracy (FTA).}
FTA is the harmonic mean of RTA (Recall of Template Accuracy) and PTA (Precision of Template Accuracy) proposed by \citet{khan2022guidelines}.
FTA has a different definition of ``correct identification'' from FGA, and we define a new notation $\hat{N}_c$ to represent the number of templates that are correctly identified by a log parser.
For FTA, one template is regarded as correctly identified if and only if these two conditions hold: (1) the parsed template's corresponding set of log messages share the same ground-truth template; (2) all the tokens of the template are the same as those of the ground-truth template.
Then, we can define PTA as $\frac{\hat{N_c}}{N_p}$ and RTA as $\frac{\hat{N_c}}{N_g}$.
And FTA can be calculated as their harmonic mean, \ie $\frac{2\times PTA\times RTA}{PTA+RTA}$.
FTA focuses more on the ability to identify concrete constant and parameter parts for a particular log message in comparison to FGA.

\vspace{-8pt}
\subsection{Evaluation Setup}
\label{sec: eval_setup}

For our evaluation, we carefully select 15 state-of-the-art log parsers from the literature.
Thirteen of them have been previously evaluated by~\citet{khan2022guidelines}.
Different from their work, we exclude LKE~\cite{fu2009LKE} from our evaluation, which involves the computation of pair-wise distances, rendering it impractical for large-scale scenarios.
These log parsers are all statistic-based, using techniques that are based on frequency (\ie LFA~\cite{nagappan2010abstracting}, LogCluster~\cite{vaarandi2015logcluster}, Logram~\cite{dai2020logram}, SLCT~\cite{vaarandi2003data}), similarity (\ie LenMa~\cite{shima2016LenMa}, LogMine~\cite{hamooni2016logmine}, LogSig~\cite{tang2011logsig}), and heuristics (\ie AEL~\cite{jiang2008abstracting}, Drain~\cite{he2017drain}, IPLoM~\cite{makanju2009clustering},  MoLFI~\cite{messaoudi2018search}, SHISO~\cite{mizutani2013SHISO},  Spell~\cite{du2016spell}).
For implementation, we directly reuse the source codes released by previous work~\cite{zhu2019tools,khan2022guidelines}.
Moreover, we have incorporated two semantic-based log parsers, namely UniParser~\cite{liu2022uniparser} and LogPPT~\cite{le2023log}, into our benchmarking study.
We implement the UniParser model following the details provided in its corresponding paper and reuse the source code of LogPPT.

In our evaluation, we apply the same preprocessing rules (\eg regular expressions) and fine-tune the parameter settings through multiple runs of each log parser.
For log parsers that exhibit variability in their parsing results due to inherent randomness, \eg MoLFI and LogPPT, we perform the evaluation five times.
By reporting the median result, we aim to mitigate potential biases arising from such randomness and present a more reliable assessment of their performance.
All experiments were conducted on a server equipped with an Intel(R) Xeon(R) Gold 6226R CPU @ 2.90GHz, 256GB RAM, and an NVIDIA GeForce GTX3090, running Ubuntu 16.04.7 LTS.

\section{Study Results}

\vspace{-2pt}
\subsection{RQ1: Differences between \nm and \olddata}
In this RQ, we aim to investigate the difference in data characteristics between \nm and \olddata.
As previously indicated in Table~\ref{tab: dataset}, \nm significantly surpasses \olddata in terms of the sizes of log messages and templates, with approximately 1,900 times more messages and 3 times more templates on average.
This substantial increase might imply a significant distinction  in feature distribution across these two dataset collections.

As discussed in Sec.~\ref{sec: motivation}, there are two pivotal characteristics inherent to log datasets: the frequency of templates and the parameter count of templates.
Specifically, the frequency of a log template refers to the number of log messages belonging to a specific log template. 
The parameter count of a log template is the number of different dynamic parts within each log template, \ie the number of \texttt{"<*>"} symbols in a log template. 
We calculate the distribution of these two characteristics for each dataset in \olddata and \nm, and plot the corresponding cumulative distribution function diagrams.
Due to space constraints, we only present three representative datasets in \nm, \ie Spark, Linux, and OpenSSH.
The figures for all 14 datasets are available at our repository~\cite{repo}.

\begin{figure}[t]
    \centering
    \includegraphics[width=\linewidth]{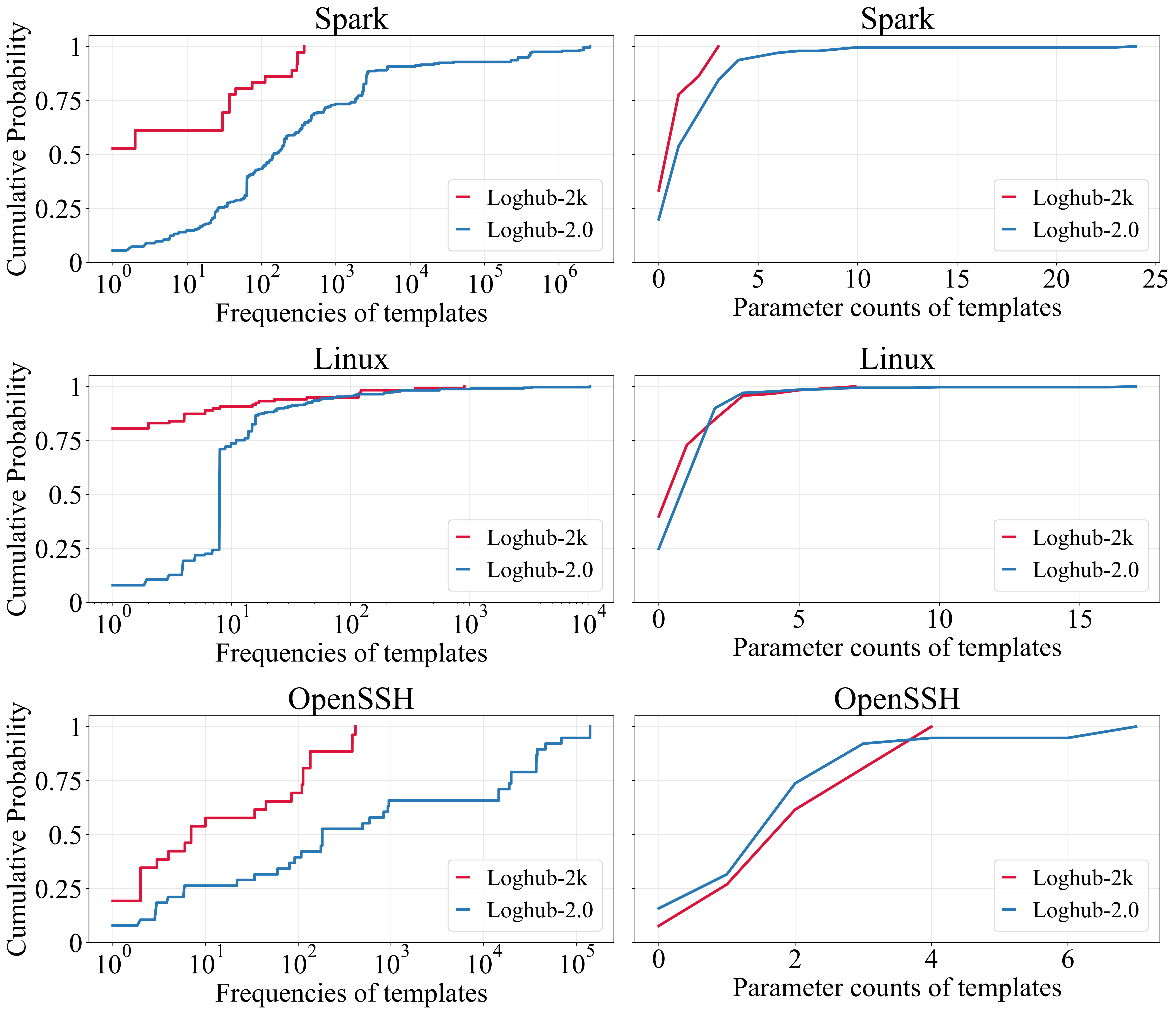}
    \caption{Distribution comparison of template frequencies and parameter counts in \nm and \olddata}
    \label{fig: rq1_result}
    \vspace{-15pt}
\end{figure}

\noindent
\textbf{The distribution of templates' frequencies}
The figures on the left-hand side of Fig.~\ref{fig: rq1_result} depict the distribution of template frequencies across three datasets.
On the one hand, \nm exhibits a broader range of template frequencies, \eg in Spark of \nm, the template frequencies range from 1 to over $10^6$, while in the \olddata, the range is narrower, ranging from 1 to around $10^3$.
On the other hand, the long-tail distribution of \nm is more pronounced than that of \olddata, indicating more imbalanced template frequencies.
For example, in the Spark dataset of \nm, only 10\% of the templates have frequencies exceeding $10^4$, yet these few templates constitute the majority of the logs.

\noindent
\textbf{The distribution of templates' parameter counts}
The three figures on the right-hand side of Fig.~\ref{fig: rq1_result} present the distribution of templates' parameter counts. 
We can observe that \nm covers a wider array of templates, each with a significantly higher number of parameters compared to those in \olddata. 
For example, the maximum number of parameters of Spark's log templates is 3 in \olddata. 
However, this number rises to 24 in the case of \nm. A similar trend is observed for both Linux and OpenSSH, suggesting that \nm has more complex log templates. 
This complexity presents a greater challenge for a log parser in accurately identifying an increased number of parameters.

\vspace{-4pt}
{
\begin{tcolorbox}[boxsep=1pt,left=2pt,right=2pt,top=3pt,bottom=2pt,width=\linewidth,colback=white!90!black,boxrule=0.2pt,colbacktitle=white!,toptitle=2pt,bottomtitle=1pt,opacitybacktitle=0,breakable]
\textbf{Finding 1.} 
The distributions of log template frequencies and parameter counts in \olddata and \nm exhibit significant differences. \nm, in particular, exhibits a more pronounced imbalance in template frequencies. Additionally, \nm contains a larger number of templates, and each has a larger parameter count on average compared to those in \olddata.
\end{tcolorbox}
}
\vspace{-4pt}

\vspace{-2pt}
\subsection{RQ2: Performance differences of log parsers on \nm and \olddata}
Given the differences in characteristics observed between \olddata and \nm in RQ1, the potential impact of such differences on the performance of log parsers remains unclear.
To address this, we apply the selected 15 log parsers to \nm and compare the performance with \olddata~\cite{zhu2019tools,khan2022guidelines} in terms of effectiveness and efficiency.
Specifically, we apply the same experimental settings (\eg preprocessing and parameter tuning) described in Sec.~\ref{sec: eval_setup} to all the evaluated log parsers.
To evaluate their effectiveness, we report metrics including GA, PA, FGA, and FTA for both \nm and \olddata. 
Additionally, we record the parsing time for each parser, which is measured from the beginning of loading log data to the completion of parsing.
Following existing work~\cite{zhu2019tools, khan2022guidelines}, we set a timeout of 12 hours. If a parser cannot finish parsing a dataset within this timeframe, we terminate the process and mark it as ``timed out''. 
Any parser that surpasses this time limit might not be suitable for practical deployment in a production environment, which often handles massive amounts of log data on a daily basis~\cite{zhu2019tools,wang2022spine,ma2018kernel}. Due to the space limitation, interested readers can refer to our repository~\cite{repo} for more detailed evaluation results.

\begin{figure*}[t]
    \centering
    \includegraphics[width=\textwidth]{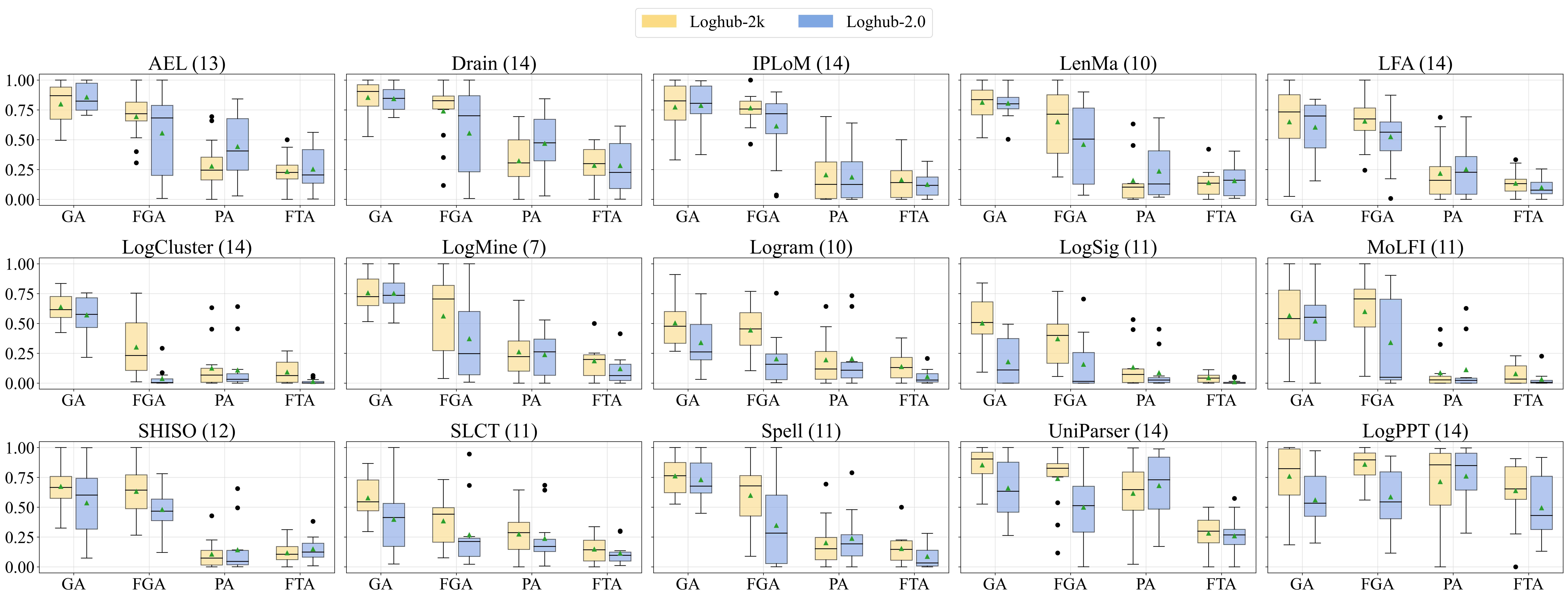}
    \captionsetup{justification=centering}
    \caption{The evaluation results of all log parsers on \olddata and \nm}
    \label{fig: rq2_result}
    \vspace{-6pt}
\end{figure*}

\subsubsection{Effectiveness}
\label{sec: rq2_effectiveness}
Fig.~\ref{fig: rq2_result} presents a box plot illustrating the effectiveness of all log parsers on \olddata and \nm. Each box encapsulates the distribution of experimental results across all datasets in terms of a specific metric. 
In addition, we also denote the number of datasets each log parser can finish processing within 12 hours in the parentheses next to the parser's name.

According to Fig.~\ref{fig: rq2_result}, we can make the following observations.
(1) For most log parsers, whether applied to \olddata or \nm, the average FGA is typically lower than GA. 
This shows that FGA is a more strict metric because it fairly treats all types of templates without considering their frequencies.
(2) When comparing GA and FGA across \olddata and \nm, we can find that the decrease of FGA in \nm is more obvious than that in \olddata. For example, for AEL, the discrepancy between the average GA and FGA on \olddata is about 0.1. However, on \nm, this value escalates to roughly 0.3.
This indicates that the template distribution in \nm is more imbalanced than \olddata, which validates our finding in RQ1.
(3) Similarly, FTA generally decreases when compared with PA, either in the \olddata or \nm, suggesting that PA is also dominated by major classes.

{
\begin{tcolorbox}[boxsep=1pt,left=2pt,right=2pt,top=3pt,bottom=2pt,width=\linewidth,colback=white!90!black,boxrule=0.2pt, colbacktitle=white!,toptitle=2pt,bottomtitle=1pt,opacitybacktitle=0,breakable]
\textbf{Finding 2.}
Message-level metrics, such as GA and PA, usually produce higher evaluation results compared to template-level metrics like FGA and FTA, due to their sensitivity to imbalanced log data. 
The differences between these metrics are more noticeable in the large-scale \nm, which displays greater imbalances.
\end{tcolorbox}
}

In addition, it is obvious that the performance of all log parsers across all metrics displays a significant difference between the \olddata and \nm. Specifically,
(1) When comparing \olddata with the more imbalanced \nm, we generally observe an increase in PA (\eg in AEL and Drain) and a slight decrease in GA for most parsers (\eg LenMA and LFA).
This is attributed to the fact that GA demands precise grouping of log messages belonging to the same templates, while PA is calculated based on the accuracy of parsing individual log messages.
The task of accurate grouping becomes more challenging within the larger \nm, leading to a noticeable increase in PA and a slight decrease in GA when using \nm.
(2) All log parsers display a significant drop in template-level metrics on \nm compared to their performance on \olddata.
For instance, Drain, which achieves the highest FGA metric on \olddata, sees its average FGA score dropping from 0.75 to approximately 0.55.
Similarly, LogPPT, though achieving the highest FTA on \olddata, experiences a reduction in its average FTA from roughly 0.64 to 0.5.
(3) Furthermore, it is noteworthy that the variances of the four metrics across different datasets for most log parsers (\eg AEL, Drain, LenMa and LogMine) have significantly increased, visually represented by the expanded range of the box plot. This implies that existing parsers struggle to achieve consistent effectiveness across different large-scale systems.

{
\begin{tcolorbox}[boxsep=1pt,left=2pt,right=2pt,top=3pt,bottom=2pt,width=\linewidth,colback=white!90!black,boxrule=0.2pt, colbacktitle=white!,toptitle=2pt,bottomtitle=1pt,opacitybacktitle=0,breakable]
\textbf{Finding 3.}
The evaluation results obtained on the \olddata do not consistently hold when the log parsers are applied to the large-scale \nm. 
On \nm, existing parsers experience a performance drop and an increase in the variance of all metrics.

\end{tcolorbox}
}

Additionally, semantic-based log parsers, such as UniParser and LogPPT, have consistently demonstrated notably higher PA and FTA scores compared to other log parsers on both \olddata and \nm datasets. 
This suggests that semantic information can facilitate accurately identifying the template of individual log messages.
However, their GA and FGA scores are generally lower than those of other log parsers. 
This can be attributed to their disregard for global information, such as statistical frequency.
As a result, these parsers are more prone to categorizing logs from the same templates into different groups, leading to lower GA and FGA scores.
Furthermore, although semantic-based log parsers have achieved commendable performance on \olddata, the performance metrics also decrease dramatically when applied to \nm. 
The primary reason is that \olddata contain too few log messages and log templates, making it easy for these models to learn the features of the entire dataset.
For instance, in LogPPT, the default number of prompts for tuning is 32, however, many datasets in \olddata have even fewer than 32 templates, resulting in LogPPT achieving near-perfect accuracy on these datasets. In contrast, since the number of log messages and log templates in \nm significantly increases, it becomes challenging for these models to generalize to more unseen log messages based on limited training samples.

{
\begin{tcolorbox}[boxsep=1pt,left=2pt,right=2pt,top=3pt,bottom=2pt,width=\linewidth,colback=white!90!black,boxrule=0.2pt, colbacktitle=white!,toptitle=2pt,bottomtitle=1pt,opacitybacktitle=0,breakable]
\textbf{Finding 4.} 
Semantic-based log parsers are more capable of parsing individual logs, evidenced by their higher PA and FTA. However, they exhibit lower grouping-related metrics, due to their neglect of global information. 
Moreover, the performance of these log parsers may decline on larger and more diverse datasets in \nm, particularly when the number of annotated samples available for training is limited.
\end{tcolorbox}
}

\subsubsection{Efficiency}
\label{sec: efficiency}
For the \olddata, all log parsers can successfully parse all 14 datasets.
However, when these log parsers are applied to the larger-scale datasets of \nm, most of them (9 out of 15) are unable to complete the parsing process for all 14 datasets within 12 hours.
Due to the space limit, we have uploaded the detailed time cost for each parser processing each dataset to our replication repository~\cite{repo}.
As depicted in Fig.~\ref{fig: rq2_result}, only six parsers (\ie Drain, IPLoM, LFA, LogCluster, LogSig, UniParser, and LogPPT) successfully complete parsing on all 14 datasets of \nm.
Certain parsers, such as LenMa and LogMine, despite demonstrating superior performance, are unable to process larger datasets efficiently.
Considering the substantial demands for log parsing throughput in real-world systems, \eg millions of logs per hour~\cite{wang2022spine}, log parsers that are unable to complete the parsing process within a reasonable timeframe (\ie 12 hours) may have limited applicability in practical scenarios.
Moreover, semantic-based log parsers like LogPPT require GPU computational resources.
When computing with a CPU, their time consumption is considerably higher compared to other efficient statistic-based log parsers like Drain.
This potentially hampers their adoption in scenarios with  resource constraints.% such as edge computing.

\vspace{-5pt}
{
\begin{tcolorbox}
[boxsep=1pt,left=2pt,right=2pt,top=3pt,bottom=2pt,width=\linewidth,colback=white!90!black,boxrule=0.2pt, colbacktitle=white!,toptitle=2pt,bottomtitle=1pt,opacitybacktitle=0,breakable]
\textbf{Finding 5.}
9 out of 15 log parsers are unable to process all 15 datasets of \nm within a reasonable 12-hour timeframe. 
Moreover, semantic-based methods typically demand more computational resources than statistic-based parsers.
\end{tcolorbox}
}
\vspace{-5pt}

\vspace{-5pt}
\subsection{RQ3: Performances of log parsers on logs with different characteristics}
Previous studies~\cite{zhu2019tools,khan2022guidelines} typically report the overall performance of log parsers on the entire full set of a dataset.
However, this may not fully characterize their effectiveness on logs with diverse characteristics, especially those that demand more attention in real-world system maintenance.
To address this limitation, our benchmarking study adopts a more granular approach by evaluating log parsers on specific logs with varying characteristics.
We primarily focus on the characteristics of template frequency and parameter count.
In specific, log templates with a lower frequency often represent rare events, hidden problems, and potential failures.
Besides, log templates with more parameters might be more informative to on-site engineers for analysis.
Thus, it is crucial to accurately parse these two types of log messages.

To this end, we first apply log parsers to parse the entire dataset.
We then select subsets of logs with different characteristics and calculate the performance on each subset. 
This approach ensures that the input data for each parser are consistent with that in RQ2, instead of merely parsing the selected subset of logs.
Due to the space limitation, we only present the results of five representative log parsers that are capable of parsing the majority of datasets in \nm.
The complete results can be found in our repository~\cite{repo}.

\begin{figure*}[t]
    \centering
    \includegraphics[width=\textwidth]{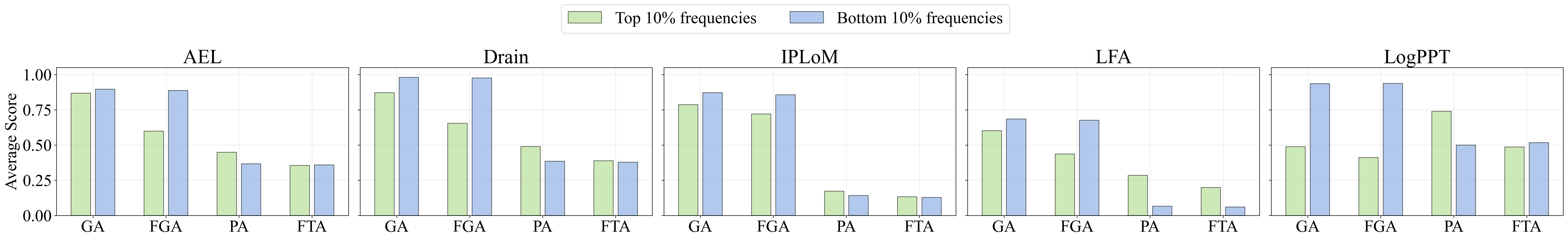}
    \caption{The evaluation results of log parsers on logs with different frequencies}
    \label{fig: rq3_frequency}
    \vspace{-6pt}
\end{figure*}
\begin{figure*}[t]
    \centering
    \includegraphics[width=\textwidth]{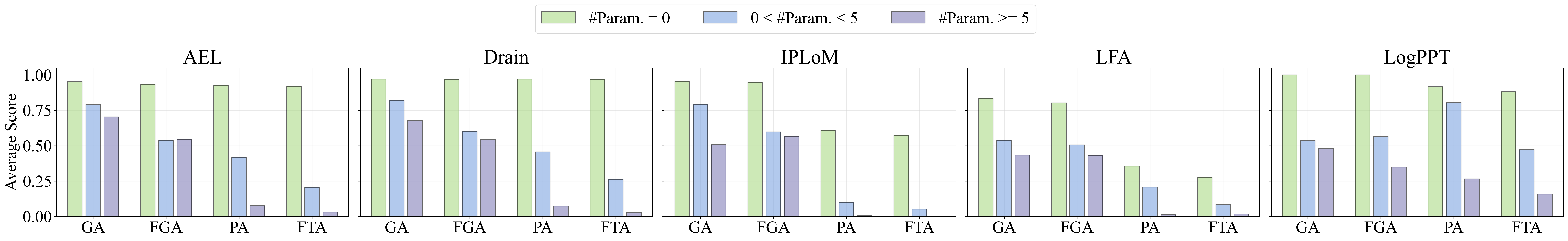}
    \caption{The evaluation results of log parsers on logs with different parameter counts}
    \label{fig: rq3_complexity}
    \vspace{-6pt}
\end{figure*}

\subsubsection{Performance with different template frequencies}
\label{sec: impact1}
As mentioned in RQ1, \nm exhibits a higher imbalance in template frequencies. 
Considering the data-driven nature of log parsers, their performance could be affected. 
Hence, we investigate the performance on template frequencies by looking into the relative infrequent and frequent logs.
In particular, we evaluate log parsers on the templates with top and bottom k\% frequencies, where k is set as 5, 10, and 20, respectively.
Then, we report the average scores of these four metrics. 
Fig.~\ref{fig: rq3_frequency} illustrates the results when k=10, while the results for k=5 and 20 can be found in our repository~\cite{repo}.

As illustrated in Fig.~\ref{fig: rq3_frequency}, all log parsers exhibit worse GA and FGA on frequent logs than on infrequent ones.
Taking LogPPT as an example, its GA and FGA scores approach 0.95 for infrequent templates, while dropping below 0.5 for frequent ones.
We explain the performance drop as follows. Considering grouping accuracy requires a parser to correctly group all logs that belong to a certain template, the grouping will become more challenging as more log messages should be included.

In contrast, all log parsers demonstrate lower average PA for infrequently occurring log templates compared to frequent ones.
This is expected for statistic-based log parsers, as the less frequency of a template provides less information and evidence (\eg count discrepancy between static and dynamic tokens) for the parser, resulting in decreased parsing accuracy.
For semantic-based log parsers, such as LogPPT, which typically require a sample of logs for training, the sampling process reduces the likelihood of selecting infrequent templates.
This, in turn, decreases the parsing accuracy of infrequent templates compared to high-frequency templates.

\vspace{-6pt}
{
\begin{tcolorbox}[boxsep=1pt,left=2pt,right=2pt,top=3pt,bottom=2pt,width=\linewidth,colback=white!90!black,boxrule=0.2pt, colbacktitle=white!,toptitle=2pt,bottomtitle=1pt,opacitybacktitle=0,breakable]
\textbf{Finding 6.}
Existing log parsers exhibit varying effectiveness when dealing with templates of different frequencies. 
They typically achieve lower GA and FGA for frequent templates, as the grouping is more challenging for templates with more log messages. 
Besides, they achieve lower PA and FTA for infrequent templates, as less evidence (\eg training data) is available to guide the accurate parsing of each log message.
\end{tcolorbox}
}
\vspace{-6pt}

\vspace{-6pt}
\subsubsection{Performance with different parameter counts}
Our study in RQ1 demonstrates that parameter counts of templates can vary in a large range (\eg 0 to 25 for the Spark dataset). 
Therefore, we also evaluate parsing effectiveness for templates with different numbers of parameters.
To achieve this, we classify the logs in each dataset within \nm into three categories based on their parameter count: logs with no parameters, logs with one to four parameter count, and logs with five or more parameters. We then utilize the same approach in Sec.~\ref{sec: impact1} to calculate the average of four metrics for each log parser on each category, respectively.

According to the results illustrated in Fig.~\ref{fig: rq3_complexity}, we can observe that all log parsers exhibit a significant decline across all four performance metrics as the parameter counts increase.
More specifically, these parsers perform exceptionally well on logs without parameters, significantly surpassing their overall performance on all logs as presented in Fig.~\ref{fig: rq2_result}.
For example, Drain achieves an average score exceeding 0.95 on all four metrics on logs without parameters, much higher than overall performance on the complete set of logs. 
When dealing with logs with more than five parameters, all log parsers exhibit notably poor performance. For example, LogPPT only attains an average FGA of 0.16, while the average FGA of other methods does not surpass 0.03. 
This suggests that despite many log parsers demonstrating relatively high performance on entire datasets, their performance on logs with more parameters remains less than satisfactory, potentially resulting in distracting parsing errors in real-world applications.

\vspace{-4pt}
{
\begin{tcolorbox}[boxsep=1pt,left=2pt,right=2pt,top=3pt,bottom=2pt,width=\linewidth,colback=white!90!black,boxrule=0.2pt, colbacktitle=white!,toptitle=2pt,bottomtitle=1pt,opacitybacktitle=0,breakable]
\textbf{Finding 7.}
Despite the high scores achieved by the log parsers on the entire datasets, their parsing effectiveness remains unsatisfactory when dealing with parameter-intensive log templates.
\end{tcolorbox}
}
\vspace{-4pt}

\vspace{-4pt}
\subsection{Summary of all research questions}
We can make the following summaries of all research questions:
(1) The proposed collection of large-scale datasets for log parsing, \nm, exhibits significantly different characteristics of log data compared to the commonly used \olddata. 
\nm presents greater challenges for existing log parsers due to its larger scale and more complex characteristics.
(2) Our evaluation results indicate that Drain is the most performant parsers that are more capable of grouping log messages, as evidenced by the highest average GA and FGA.  
On the other hand, semantic-based methods (\eg UniParser and LogPPT) exhibit stronger abilities in distinguishing each token as either constant or dynamic parts. 
However, these methods compromise their effectiveness in grouping log messages with the same templates. This is because classification errors in tokens can easily lead to incomplete groups.
(3) Despite the encouraging results shown in the \olddata, the parsing performance remains unsatisfactory when applied to \nm. This is particularly noticeable when parsing infrequent logs and parameter-intensive logs.
(4) Moreover, the efficiency of the majority of log parsers fails to meet the demands of large-scale application scenarios.

\vspace{-7pt}
\section{Discussion}

\subsection{Implications}

Based on our findings, we have identified the following implications, which we believe could benefit future research on log parsing.

\noindent
\textbf{Consider both levels of metrics in combination.}
While most existing tasks utilize message-level metrics such as GA and PA to assess performance, these measures are often dominated by log templates with high frequencies in large-scale application scenarios, thereby yielding higher scores.
In contrast, template-level metrics are resistant to the imbalanced frequencies of templates and thus can accurately reflect the parsing performance on datasets with diverse template distributions.
Hence, these two types of metrics may be contemplated in conjunction, and one can be prioritized over the other based on specific requirements.
For instance, if the focus is more on the parsing accuracy of frequent log templates and one can tolerate errors in infrequent templates, then message-level metrics are more appropriate, and vice versa.

\noindent
\textbf{Evaluate the performance across logs with different characteristics.}
Although certain log parsers have exhibited high overall performance on specific datasets, their parsing performance is still lacking when handling infrequent and parameter-intensive log templates.
Considering the importance of these logs, as underscored in Section~\ref{sec: motivation}, it is crucial to concentrate specifically on performance within these logs.
The evaluation protocol we propose can unearth the performance of log parsers on these log templates more comprehensively.
Consequently, future work should pay attention to this aspect when designing new log parsers, thereby enhancing their applicability in real-world scenarios.

\noindent
\textbf{Place greater emphasis on efficiency.}
As discussed in Sec.~\ref{sec: efficiency}, many existing log parsers fail to meet the performance requirements of large-scale application scenarios, a fact not represented in \olddata. Considering the large volume of logs in practical settings, it is imperative that future log parsers are designed to meet the performance demands of specific applications.

\noindent
\textbf{Try to combine semantic and statistical information.}
According to finding 4, semantic-based log parsers possess superior capabilities in distinguishing parameters from templates, which also substantiates the significance of semantic information in the process of log parsing.
However, their grouping abilities are compromised due to the neglect of global information.
This is inevitable, given that these log parsers exclusively process each log message in isolation.
A potential avenue for future research could involve the combination of semantic and statistical information in logs, thereby simultaneously enhancing the parsing and grouping capabilities.

\vspace{-3pt}
\subsection{Threats to Validity}

\noindent
\textbf{Annotation errors}
The primary threat of this study is the potential annotation errors in \nm, which is inevitable without the source code. To mitigate this issue as much as possible, we have designed a stringent annotation framework with a team comprising five members with significant experience in log analysis research.

\noindent
\textbf{Limited log parsers}
The selection of log parsers is limited, as not all existing log parsers are open-sourced due to industry confidentiality reasons~\cite{wang2022spine}.
Nevertheless, the selected parsers have included state-of-the-art log parsers published at top-tier conferences and covered all existing categories of technology.
Furthermore, we have made our dataset available and implemented our benchmark protocol in a unified and user-friendly manner. 
This allows for the easy comparison of additional log parsers with existing ones.

\noindent
\textbf{Implementation and settings}
To mitigate the bias of implementation and settings, we have adopted the source code of several log parsers from the widely-used benchmark~\cite{zhu2019tools,le2023log}. For the newly incorporated log parsers, we have either used the source code provided by the original authors or carefully replicated them to ensure the fidelity of the results. Additionally, we have tuned the parameters of each log parser to optimize the results.

\vspace{-2pt}
\section{Conclusion}

In this paper, we conduct a more rigorous and practical large-scale evaluation for log parsing techniques.
We propose a log template annotation framework that ensures both efficiency and accuracy, and have annotated a new collection of large-scale datasets for log parsing, which more accurately reflects the scale and distribution of log data in real-world situations.
Our proposed benchmarking protocol, inclusive of a new template-level metric and an evaluation of the performance of log parsers on logs with varying characteristics, offers a more comprehensive and in-depth analysis of log parsers' performance.
Furthermore, our re-evaluation of selected log parsers using \nm uncovers valuable findings of the limitations of existing log parsers and benchmarks.
We believe that our work, together with the open-source dataset \nm and benchmark, could benefit future research in the field of log analysis.

\vspace{-2pt}
\section*{Acknowledgment}

The work described in this paper was supported by the Research Grants Council of the Hong Kong Special Administrative Region, China (No. CUHK 14206921 of the General Research Fund). We extend our sincere gratitude to the anonymous reviewers for their constructive feedback.

\balance
\bibliographystyle{ACM-Reference-Format}
\bibliography{ISSTA24}
\end{document}